\documentclass[12pt,a4paper]{article}
\usepackage[latin9]{inputenc}
\usepackage{amsmath}
\usepackage{amssymb}
\usepackage{esint}

\makeatletter

\pdfpageheight\paperheight
\pdfpagewidth\paperwidth

\pdfoutput=1 

\usepackage{jheppub}

\usepackage{etoolbox}
    
    \patchcmd{\maketitle}{\@fpheader}{}{}{}

\usepackage{amsfonts}
\usepackage{bbm}
\usepackage{amssymb}
\usepackage{amsmath}

\usepackage{graphicx,color}
\usepackage{slashed}
\usepackage{young}
\usepackage[vcentermath]{youngtab}

\makeatletter

\newcommand{\ep}{\epsilon}

\newcommand{\be}{\begin{equation}}
\newcommand{\ee}{\end{equation}}
\newcommand{\bea}{\begin{eqnarray}}
\newcommand{\eea}{\end{eqnarray}}

\title{\boldmath Higher Spin Conformal Geometry in Three Dimensions and Prepotentials for Higher Spin Gauge Fields}

\author{Marc Henneaux,}
\author{Sergio H\"ortner,}
\author{and Amaury Leonard}

\affiliation{Universit\'{e} Libre de Bruxelles and International Solvay Institutes,
ULB Campus Plaine C.P.231, B-1050 Bruxelles, Belgium.}

\abstract{We study systematically the conformal geometry of higher spin bosonic gauge fields in three spacetime dimensions. We recall the definition of the Cotton tensor for higher spins and establish a number of  its properties that turn out to be key in solving in terms of prepotentials the constraint equations of the Hamiltonian ($3+1$) formulation of four-dimensional higher spin gauge fields.       The prepotentials are shown to exhibit higher spin conformal symmetry.  Just as for spins 1 and 2, they provide a remarkably simple, manifestly duality invariant formulation of the theory.   While the higher spin conformal geometry is developed for arbitrary bosonic spin, we explicitly perform the Hamiltonian analysis and derive the solution of the constraints only in the illustrative case of spin $3$. In a separate publication, the Hamiltonian analysis in terms of prepotentials is extended to all bosonic higher spins using the conformal tools of this paper, and the same emergence of higher spin conformal symmetry is confirmed.}

\makeatother

\begin{document}
\maketitle \flushbottom

\section{Introduction}
\setcounter{equation}{0}

Three dimensional space is well known to be peculiar from the point of view of conformal geometry (see e.g. \cite{Eisenhart}). Indeed, while the Weyl tensor controls the conformal geometry in dimensions $D \geq 4$, this tensor turns out to identically vanish in $D=3$.  What plays the role of the Weyl tensor is then the Cotton tensor, which depends on the metric and its derivatives up to third order and which, in the appropriate dual representation, has the following properties:
\begin{enumerate}
\item It is a rank-two symmetric tensor;
\item It is traceless;
\item It is divergence-free:
\item It is invariant under Weyl rescalings;
\item It controls conformal invariance, in the sense that a necessary and sufficient condition for a metric to be conformally flat is that its Cotton tensor vanishes.
\end{enumerate}
The Cotton tensor appears in three-dimensional topologically massive gravity \cite{Deser:1981wh}.

In the case of linearized gravity around a flat background with $g_{ij} = \delta_{ij} + h_{ij}$, these properties for the Cotton tensor 
\begin{eqnarray}
B^{ij}[h] &=&  \frac{1}{4} \ep^{imn} \left(\partial^j \partial_n \partial^s h_{sm} - \triangle \partial_n h^j_{\; \;  m} \right) \nonumber \\ && + \frac{1}{4} \ep^{jmn} \left(\partial^i \partial_n \partial^s h_{sm} - \triangle \partial_n h^i_{\; \;  m} \right) 
 \label{DefD01}
\end{eqnarray}
read
\begin{enumerate}
\item  Symmetry: 
\be  
B^{ij} = B^{ji}, \label{PropD0-1}
\ee
\item Tracelessness:
\be
B^{ij}\delta_{ij} = 0, \label{PropD00}
\ee
\item Transverseness: 
\be \partial_i B^{ij} = 0, \label{PropD01}
\ee
\item Gauge invariance:
\be 
\delta B^{ij} = 0
\; \; \; \; \; \; \hbox{ for 
$\delta h_{ij} = 2 \partial_{(i}\xi_{j)} + \lambda \delta_{ij}$}, \label{PropD02}
\ee
\item Conformal flatness:
\be
B^{ij}[h] = 0 \; \Leftrightarrow \; h_{ij} = 2 \partial_{(i}\xi_{j)} + \lambda \delta_{ij}
\ee

for some $\xi_j$ and $\lambda$. 
\end{enumerate}
 We have assumed Euclidean signature, but similar formulas with $\delta_{ij}$ replaced by the Minkowskian metric of course hold in the case of Minkowskian signature.

Furthermore, one can show that :
\begin{enumerate}
\setcounter{enumi}{5}
\item Any tensor $B^{ij}$ that fulfills (\ref{PropD0-1}), (\ref{PropD00}) and (\ref{PropD01}) can be written as in (\ref{DefD01}), i.e., is the Cotton tensor of some $h_{ij}$.  
\end{enumerate}
This  property turns out to be crucial when analyzing the constraints of the spin-2 theory in four dimensions  as we shall recall below.

Conformal higher spin gauge fields have received a sustained interest over the years  \cite{Fradkin:1985am,Pope:1989vj,Fradkin:1989md,Fradkin:1989xt,Vasiliev:2001zy,Segal:2002gd,Shaynkman:2004vu,Vasiliev:2007yc,Metsaev:2007rw,Marnelius:2008er,Metsaev:2009ym,Vasiliev:2009ck,Florakis:2014kfa,Nutma:2014pua}.  The gauge symmetries of a conformal bosonic higher spin gauge field of spin $s$ read, in the free limit,
\begin{equation} \delta h_{i_1 \cdots i_s} = s \partial_{(i_1} \xi_{i_2 \cdots i_s)} + \frac{s(s-1)}{2}\delta_{(i_1 i_2} \lambda_{i_3 \cdots i_s)} \label{GaugeTransf}
\end{equation}
The gauge transformations parametrized by $ \xi_{i_2 \cdots i_s}$ reduce to the Maxwell gauge transformations $\partial_i \xi$ when $s=1$ and to linearized diffeomorphisms when $s=2$.   They will be called here ``spin-$s$ diffeomorphisms" (or ``higher spin  diffeomorphisms" when we do not need to specify the explicit value of $s>2$).  The gauge transformations parametrized by $\lambda_{i_3 \cdots i_s}$ are absent for $s=1$ and reduce to linearized Weyl rescalings $\delta_{i_1 i_2} \lambda $ when $s =2$.  They will be called here ``spin-$s$ Weyl transformations" (or ``higher spin Weyl transformations", or ``higher spin conformal transformations"). 
Similarly, a fermionic conformal higher spin field of spin $s + \frac12$ is described by a tensor spinor $\Sigma_{i_1 \cdots i_s}$ with gauge symmetries
\begin{equation} \delta \Sigma_{i_1 \cdots i_s} = s \partial_{(i_1} \mu_{i_2 \cdots i_s)} + s \gamma_{(i_1} \eta_{i_2 \cdots i_s)} . \label{GaugeFermionic}
\end{equation}
Both sets of gauge transformations are reducible since $\delta h_{i_1 \cdots i_s}= 0$ for 
$$\xi_{i_2 \cdots i_s} = \frac{(s-1)(s-2)}{2} \delta_{(i_2 i_3}\psi_{i_4 \cdots i_s)}, \; \; \; \lambda_{i_3 \cdots i_s} = - (s-2)\partial_{(i_3}\psi_{i_4 \cdots i_s)},$$ while $\delta \Sigma_{i_1 \cdots i_s} = 0$ for $\mu_{i_2 \cdots i_s} = \gamma_{(i_2} \zeta_{i_3 \cdots i_s)}$ and $\eta_{i_2 \cdots i_s} = - \partial_{(i_2} \zeta_{i_3 \cdots i_s)}$.  One could use this redundancy to impose trace conditions on the gauge parameters but this will not be done here.  

We shall from now on focus on the bosonic (integer spin) case. Building the higher spin conformal geometry amounts to constructing a complete set of invariants under (\ref{GaugeTransf}) out of the fields and their derivatives.  It turns out that the case of spacetime dimension $D \geq 4$ is again rather direct and uneventful, because a straightforward generalization of the Weyl tensor provides the solution.  In dimension $D=3$, however, the Weyl tensors for higher spins vanish identically, just as for spin-2.  One must introduce the Cotton tensor \cite{Damour:1987vm,Pope:1989vj}.  The properties of the Cotton tensor were thoroughly explored in the profound work \cite{Damour:1987vm} for spin $s=3$.  In particular, the analog of the properties 1-5 for $s=3$ were explicitly demonstrated.  The purpose of this paper is to extend the analysis of \cite{Damour:1987vm} to higher spins, building on the previous study \cite{Pope:1989vj}.  We also prove the generalization of property 6 for all spins. 

One motivation for undertaking the present study is that the Cotton tensor plays a central role in some recent higher spin gauge models in three dimensions \cite{Bergshoeff:2009tb,Bergshoeff:2011pm,Nilsson:2013tva,Nilsson:2015pua}.    Another motivation comes from $SO(2)$ electric-magnetic duality invariance, as we now discuss.

It has been recognized some time ago that  the free Maxwell action in four spacetime dimensions is invariant under $SO(2)$ rotations in the two-dimensional internal plane of the electric and magnetic fieds,
\begin{eqnarray} && {\mathcal E}^i \longrightarrow \cos \alpha \, {\mathcal E}^i  - \sin \alpha \, {\mathcal B}^i \nonumber \\
&& {\mathcal B}^i \longrightarrow \sin \alpha \, {\mathcal E}^i  + \cos \alpha \, {\mathcal B}^i \nonumber
\end{eqnarray}
It was indeed shown in the pioneering work  \cite{Deser:1976iy} that, contrary to common belief, this symmetry is not just an on-shell symmetry leaving the equations of motion invariant, but is a genuine off-shell symmetry of the action $S^{\textrm{Maxwell}}[A_\mu]$ once appropriately extended to the vector potential $A_\mu$,  which is the dynamical variable in the action principle. In the original, single vector potential formulation, the duality transformations of the vector potential are non-local in space, but locality can be achieved by going to the Hamiltonian formalism and solving Gauss' constraint through the introduction of a second vector potential \cite{Deser:1976iy}.  In the formulation with two potentials, $SO(2)$ electric-magnetic duality invariance of the action is manifest and amounts to rotations in the internal plane of the two vector potentials \cite{Deser:1976iy,Deser:1997mz}.  These two vector potentials $A^a_i$ ($ a= 1,2$) are both gauge invariant under (\ref{GaugeTransf}).

The analysis can be extended to spin 2. Electric-magnetic $SO(2)$-rotations in the internal plane spanned by the linearized Riemann tensor and its dual are symmetries of the free spin-2 theory, not only of the equations of motion, but also of the action itself \cite{Henneaux:2004jw}(see also \cite{Deser:2004xt,Julia:2005ze} for a different approach and  the extension to the cosmological case,  respectively).  The action can be written in a manifestly duality invariant form by solving the constraints \cite{Henneaux:2004jw}, as for  spin 1. This step requires the introduction of one potential for the spatial metric through the resolution of the so-called Hamiltonian constraint and one potential for its conjugate momentum through the resolution of the so-called momentum constraint.  Since the metric and its momentum are themselves already potentials for the gauge invariant (linearized) Riemann tensor, one talks of ``prepotentials". The two prepotentials $Z^a_{ij}$ are rank-2 symmetric tensors and $SO(2)$ duality transformations are simply rotations in the internal plane of the prepotentials. 
\begin{eqnarray} &&Z^1_{ij} \longrightarrow \cos \alpha \, Z^1_{ij} - \sin \alpha \, Z^2_{ij} \nonumber \\
&& Z^2_{ij}\longrightarrow \sin \alpha \, Z^1_{ij}  + \cos \alpha \, Z^2_{ij} \nonumber
\end{eqnarray}
An intriguing feature of the prepotentials is that they are both invariant under the gauge symmetries (\ref{GaugeTransf}) of conformal spin 2,
\be
\delta Z^a_{ij} = 2 \partial_{(i} \eta^a_{j)} + \delta_{ij} \mu^a
\ee
The emergence of (linearized) diffeomorphisms {\em and} Weyl rescalings is somewhat unexpected but turns out to be crucial in the investigation of the theory and the understanding of its structure \cite{Bunster:2012km}.  This intriguing feature arises also for gravity in higher dimensions \cite{Bunster:2013oaa} where the prepotentials are now tensors with different Young tableau symmetries, and  for the fermionic spin 1/2 and spin 3/2 massless fields in four dimensions, where the resolution of the constraints introduce also prepotentials with the symmetries (\ref{GaugeFermionic}) of conformal fermionic gauge fields \cite{Bunster:2012jp,Bunster:2014fca}.

These results have led  in \cite{Bunster:2014fca} to the conjecture that not only for spins 1, 3/2, 2 and 5/2 does the resolution of the constraints of the Hamiltonian formulation lead to prepotentials with the gauge symmetries of conformal gauge fields with respective spins 1, 3/2, 2 and 5/2, but that the same somewhat puzzling property holds for all higher spins.   We establish this conjecture here in the spin-$3$ case, which exhibits already the new features characteristic of higher spins.  We consider four dimensions, where the dynamical fields are three-dimensional tensors and the prepotentials completely symmetric tensors (described by single-row Young tableaux).  We find then that the validity of the conjecture is in fact a direct consequence of our  analysis of higher spin conformal geometry.  We also show that the prepotential formulation is automatically manifestly duality invariant for all spins.  While manifestly duality invariant, it is not, however, manifestly Lorentz-invariant (although it is of course Lorentz-invariant). As argued in \cite{Bunster:2012hm}, this lack of manifest spacetime covariance of the manifestly duality-invariant formulations might be the signal that the duality symmetries are more fundamental than the spacetime symmetries, which would be emergent symmetries, in line with the idea that spacetime itself is an emergent concept. Using the geometric tools developed in this paper, we show in a separate publication \cite{HHL0} that the Hamiltonian formulation of spins $>3$ exhibit the same features.

Our work is organized as follows.  In Section {\bf \ref{Sec:Riemann}}, the definition of the analog of the Riemann tensor, which contains $s$ derivatives of the spin-$s$ field \cite{deWit:1979pe}, is recalled and its main properties are reviewed. The Weyl tensor, defined as the tracefree part of the Riemann tensor, contains therefore also $s$ derivatives of the spin-$s$ field and is recalled to control the conformal geometry in spacetime dimensions $D \geq 4$. It identically vanishes in dimensions $D = 3$ for which new tools are needed and  to which we then exclusively turn (Sections {\bf \ref{Cotton}, \ref{Sec:4}, \ref{Sec:Crucial}}).  Equipped with the appropriate mathematical apparatus developed for general bosonic spins, we then discuss the Hamiltonian formalism for a spin-$3$ gauge field and show how the higher spin conformal techniques enable one to solve the constraints in terms of prepotentials that enjoy remarkable symmetries (Section {\bf \ref{Sec:Prepo}}).  In Section {\bf \ref{Sec:Duality}}, we rewrite the action in terms of the prepotentials and establish manifest $SO(2)$ electric-magnetic duality invariance.  Section {\bf \ref{Conclusions}} is devoted to concluding comments.  Finally, two appendices give our conventions and provide further cohomological insight into the higher spin Weyl symmetry.

\section{Riemann  tensor}
\label{Sec:Riemann}
\setcounter{equation}{0}
\subsection{Definition and Bianchi identity}
We first recall how to construct invariants under the spin-$s$ diffeomorphisms
\begin{equation} \delta h_{i_1 \cdots i_s} = s \partial_{(i_1} \xi_{i_2 \cdots i_s)}  \label{GaugeDiffSpinS}
\end{equation}
(without trace constraints on the gauge parameters).
This question was investigated in \cite{deWit:1979pe}, where it is shown that the relevant Riemann tensor involved $s$ derivatives of the spin-$s$ field. Explicitly, the Riemann tensor is defined by
\be
R_{i_1 j_1 i_2 j_2 \cdots i_s j_s}[h] =  2^s \partial_{[j_1\vert} \partial_{ [ j_2\vert} \cdots \partial_{[j_s\vert } h_{i_1] \vert  i_2] \vert  \cdots i_s]} 
\ee
where the antisymmetrizations are to be carried on each pair of indices $(i_k, j_k)$ ($k = 1, \cdots, s$), so that the Riemann tensor has the Young symmetry
\be
\overbrace{\yng(10,10)}^{\text{$s$ boxes}} \, . \label{YoungRiemann}
\ee
The Riemann tensor is invariant under spin-$s$ diffeomorphisms and fulfills the Bianchi identity
\be
\partial_{[k_1} R_{i_1 j_1] i_2 j_2 \cdots i_s j_s} = 0.
\ee

It will be sometimes useful to adopt an index-free notation, in order to emphasize the concepts. To that end, we shall use the nilpotent differential operators $d_{(s)}$ of order $s+1$ introduced in \cite{DuboisViolette:1999rd,DuboisViolette:2001jk},
\be
d_{(s)}^{s+1} = 0, \label{NilpotencyOrderS}
\ee
acting on ``well-filled" mixed symmetry tensors with $s$ columns (some of which can be empty) (see also \cite{Coho,XBNB} for a general discussion of cohomological techniques adapted to tensor fields with arbitrary Young symmetry). 
In these notations, the gauge transformations (\ref{GaugeDiffSpinS}) read $\delta h = d_{(s)} \xi$ and the definition of the Riemann tensor is simply $R = d_{(s)}^s h$.  The spin-$s$ diffeomorphism invariance of the Riemann tensor and the Bianchi identity both follow from (\ref{NilpotencyOrderS}). 

Furthermore, the cohomological results of \cite{DuboisViolette:1999rd,DuboisViolette:2001jk} imply:
(i) that any tensor of Young symmetry type (\ref{YoungRiemann}) that fulfills $d_{(s)} R=0$ can be written as $R = d_{(s)}^s h$ for some completely symmetric field $h_{i_1 \cdots i_s}$; and (ii) a necessary and sufficient condition for $h_{i_1 \cdots i_s}$ to be pure gauge, $h = d_{(s)} \xi$, is that its Riemann tensor vanishes.

One can prove, in fact, the equivalent result that any  function of the spin-$s$ field and its derivatives that is invariant under spin-$s$ diffeomorphisms is a function of the curvature components and their derivatives only.  A direct proof is given in  the lucid work \cite{Bekaert:2005ka}.  The equivalence of the two statements is given in Appendix {\bf {\ref{App:Complete}}}. Hence, the curvature tensor completely captures spin-$s$ gauge invariance.

The Riemann tensor is not invariant under the spin-$s$ Weyl transformations.  Rather, under these transformations, it transforms as 
\be
\delta R _{i_1 j_1 i_2 j_2 \cdots i_s j_s}= 2^s \frac{s(s-1)}{2} \Pi \left( \partial_{j_1 \cdots j_s} \delta_{(i_1 i_2} \lambda_{i_3  \cdots i_s)} \right),
\ee
where the projection operator $\Pi$ carries the antisymmetrizations within each pair of indices $(i_k, j_k)$.  The variation  of the Riemann tensor is clearly pure trace since all its terms contain a $\delta_{mn}$-factor with a pair of indices in $(i_1, j_1, i_2, j_2, \cdots, i_s, j_s)$. Therefore, the Weyl tensor, which is the trace-free part of the Riemann tensor, is Weyl invariant.  As the Riemann tensor, it contains  $s$ derivatives of the spin-$s$ field. In dimension $D \geq 4$, the Weyl tensor vanishes if and only if the spin-$s$ field is  pure gauge taking into account all the gauge symmetries of conformal spin $s$, i.e. $h_{i_1 i_2 \cdots i_s} = \partial_{(i_1} \xi_{i_2 \cdots i_s)} + \delta_{(i_1 i_2} \lambda_{i_3 \cdots i_s)}$.   This is well known for $s=2$ and was established in \cite{Damour:1987vm} for $s=3$.   The demonstration proceeds along the same lines for $s>3$.

Less direct is the case of dimension $D=3$ because the Weyl tensor vanishes identically\footnote{
The vanishing of the Weyl tensor in three dimensions is a well-known fact.  It is a direct consequence of the identities valid in three dimensions
\begin{eqnarray}
&& R_{i_1 j_1 i_2 j_2 \cdots i_s j_s} \nonumber \\
&& = \frac14\epsilon_{i_1 j_1 k_1} \epsilon^{k_1 m_1 n_1} \epsilon_{i_2 j_2 k_2} \epsilon^{k_2 m_2 n_2} R_{m_1 n_1 m_2 n_2 \cdots m_s n_s} \nonumber \\
&& = \frac14\epsilon_{i_1 j_1 k_1}  \epsilon_{i_2 j_2 k_2} \epsilon^{k_1 m_1 n_1} \epsilon^{k_2 m_2 n_2} R_{m_1 n_1 m_2 n_2 \cdots m_s n_s} \nonumber \\
&& = \frac12 \epsilon_{i_1 j_1 k_1}  \epsilon_{i_2 j_2 k_2} \delta^{k_1 k_2} R_{[2] m_3 n_3  \cdots m_s n_s}  \nonumber \\
&& \hspace{.4cm} - \epsilon_{i_1 j_1 k_1}  \epsilon_{i_2 j_2 k_2} R^{\;\; \; k_1 k_2}_{ [1]\; \; \; \; \; \; \;   m_3 n_3  \cdots m_s n_s}  \nonumber
\end{eqnarray}
showing that the Riemann tensor is entirely expressible in terms of its trace $R_{[1] k_1 k_2 \vert k_3 j_3  \cdots k_s j_s} =   \delta^{j_1 j_2} R_{k_1 j_1 k_2 j_2 \cdots k_s j_s}$ and thus that its traceless part is zero.  Here, $R_{ [2] m_3 n_3  \cdots m_s n_s}$ is the double trace $ \delta^{k_1 k_2} R_{[1] k_1 k_2 \vert m_3 n_3  \cdots m_s n_s}$.}.  To control the conformal geometry, one needs in this case the Cotton tensor, which contains $2s -1$ derivatives of the spin-$s$ field \cite{Damour:1987vm,Pope:1989vj}.

\subsection{Einstein tensor}
\label{SubSec:Einstein}
Before moving to the Cotton tensor, we introduce the Einstein tensor, which can be defined in three dimensions as
\be
G_{k_1 \cdots k_s} = \left( \frac12 \right)^s \varepsilon_{k_1 i_1 j_1} \varepsilon_{k_2 i_2 j_2} \cdots
\varepsilon_{k_s i_s j_s} R^{i_1 j_1 i_2 j_2 \cdots i_s j_s} \label{DefG}
\ee
It is completely symmetric and equivalent to the Riemann tensor, since the defining relation (\ref{DefG}) can be inverted to give
\be
R_{i_1 j_1 i_2 j_2 \cdots i_s j_s} = \varepsilon_{k_1 i_1 j_1} \varepsilon_{k_2 i_2 j_2} \cdots
\varepsilon_{k_s i_s j_s} G^{k_1 k_2\cdots k_s}
\ee
The Einstein tensor is the dual of the Riemann tensor on all pairs of indices.

The Einstein tensor fulfills the ``contracted Bianchi identity"
\be
\partial_{k_1} G^{k_1 \cdots k_s} = 0 \label{ContBianchi}
\ee
This is also true for its successive traces
\begin{eqnarray}
&& \bar{G}^{i_1 \cdots i_{s-2}} \equiv G_{[1]}^{i_1 \cdots i_{s-2}} = \delta_{i_{s-1}i_{s}} G^{i_1 \cdots i_{s-2} i_{s-1} i_s},\\
 && \bar{\bar G}^{i_1 \cdots i_{s-4}} \equiv G_{[2]}^{i_1 \cdots i_{s-4}} = \delta_{i_{s-3} i_{s-2}} \bar{G}^{i_1 \cdots i_{s-4} i_{s-3} i_{s-2}}, \hspace{1cm}
\end{eqnarray}
etc ($n = 0, \cdots \left[\frac{s}{2}\right]$, with $G_{[0]} \equiv G$), which obey
\be
\partial_{i_1} \bar{G}_{[n]}^{i_1 \cdots i_{s-2n} } = 0, \; \; \; n = 0, \cdots \left[\frac{s}{2}\right] \, . \label{SuccBianchi}
\ee
The operation of taking one trace is denoted by one bar, but multiple traces are also indicated with the subscript $[n]$ rather than multiple bars, where $n$ is the number of traces being taken.  The maximum number of traces that can be taken is equal to the integer part $\left[\frac{s}{2}\right]$ of $\frac{s}{2}$.

Just as the ``contracted Bianchi identity'' (\ref{ContBianchi}) is equivalent to $d_{(s)} R = 0$ or $d_{(s)} \, ^*G = 0$, where $^*$ denotes here the dual on all indices, the successive identities (\ref{SuccBianchi}) can be written as $d_{(s-2)} \, ^*\bar{G} = 0$, $d_{(s-4)} \, ^*\bar{\bar{G}} = 0$, etc, where $^*G_{[n]}$ are the duals of the successive traces $G_{[n]}$, which are tensors of Young symmetry type 
$$^*G_{[n]} <> \overbrace{\yng(5,5)}^{(s-2n) \text{ boxes}},$$
etc.  These identities can also be directly verified from the expression of the Einstein tensor in terms of the derivatives of the fields, 
\begin{eqnarray}
&& G^{k_1 \cdots k_s} =  \varepsilon^{k_1 i_1 j_1} \varepsilon^{k_2 i_2 j_2} \cdots
\varepsilon^{k_s i_s j_s} \nonumber \\
&& \hspace{3cm} \partial_{i_1} \partial_{i_2} \cdots \partial_{i_s} h_{ j_1 j_2 \cdots  j_s} 
\end{eqnarray}
 which implies by contraction that $^*\bar{G} = d^{s-2}_{(s-2)} \Psi$, $^*\bar{\bar{G}} = d^{s-4}_{(s-4)} \Xi$, etc, for some completely symmetric tensors $\Psi$ with $s-2$ indices, 
$$\Psi <> \overbrace{\yng(8)}^{(s-2) \text{ boxes}},$$
$\Xi$ with $s-4$ indices,
$$\Xi<> \overbrace{\yng(6)}^{(s-4) \text{ boxes}},$$
etc, which respectively depend on $2$, $4$, etc derivatives of the spin-$s$ field. 
It follows in particular from these considerations that the equation  $\bar{G} = 0$ is equivalent to $\Psi = d_{(s-2)} \Pi$, with $$\Pi <> \overbrace{\yng(7)}^{(s-3) \text{ boxes}}.$$ 

Since the Einstein tensor is equivalent to the Riemann tensor, it fully captures in the same manner the spin $s$ gauge invariance:  any  local function of the $h$'s that is invariant under the spin-$s$ gauge transformations can be expressed as a function of $G$ and its successive derivatives.  Furthermore a necessary and sufficient condition for $h$ to be pure spin-$s$ gauge is that its  Einstein tensor vanishes.  

\section{Schouten and  Cotton tensors}
\label{Cotton}
\setcounter{equation}{0}

\subsection{Schouten tensor} 
Under a spin-$s$ Weyl transformation,  the Einstein tensor transforms as
\be 
\delta G^{i_1 \cdots i_s} = \frac{s(s-1)}{2}\left(- \partial^{(i_1} \partial^{i_2} \mu^{i_3 \cdots i_s)} +  \delta^{(i_1 i_2} \triangle \mu^{i_3 \cdots i_s)}\right) \label{WeylTransG}
\ee
where 
\be 
\mu^{i_3 \cdots i_s} =   \varepsilon^{i_3 j_3 k_3} \cdots
\varepsilon^{i_s j_s k_s}  \partial_{j_3  \cdots j_s} \lambda_{k_3  \cdots k_s} \label{FormOfMu}
\ee
The tensor $\mu^{i_3 \cdots i_s}$ fulfills $\partial_{\mu_1} \mu^{i_3 \cdots i_s} = 0$ and conversely, any tensor that fulfills that equation can be written as in (\ref{FormOfMu}) with $\lambda_{k_3  \cdots k_s}$ completely symmetric.  This follows again from the cohomological theorems of \cite{DuboisViolette:1999rd,DuboisViolette:2001jk} applied now to the differential operator $d_{(s-2)}$ defined in the space of tensors with $s-2$ columns, which fulfills $d_{(s-2)}^{s-1} = 0$.

The Schouten tensor $S^{i_1 \cdots i_s}$ is then defined through
\begin{eqnarray}
S^{i_1 \cdots i_s} &=& G^{i_1 \cdots i_s} \nonumber \\
&& + \sum_{n=1}^{[\frac{s}{2}]} a_n  \delta^{(i_1 i_2}  \cdots \delta^{i_{2n-1} i_{2n}} G_{[n]}^{i_{2n+1}  \cdots i_s)} \hspace{.5cm}\label{DefSchouten}
\end{eqnarray}
where the terms added to  $ G^{i_1 \cdots i_s}$ to define $S^{i_1 \cdots i_s}$ involve the successive higher traces of $ G^{i_1 \cdots i_s}$ and are recursively adjusted in such a way that the Schouten tensor fulfills the crucial  property of  transforming as
\be
\delta S_{i_1 i_2 \cdots i_s} = -\frac{s(s-1)}{2}\partial_{(i_1} \partial_{i_2} \nu_{i_3 \cdots i_s)}  \label{TransSchouten}
\ee
under spin-$s$ Weyl transformations,  where $\nu$ is related to $\mu$ as follows 
\begin{eqnarray}
\nu^{i_3 \cdots i_s} &=&  \mu^{i_3 \cdots i_s}  \nonumber \\  
&&+  \sum_{n=1}^{[\frac{s}{2}]-1} b_n  \delta^{(i_3 i_4}  \cdots \delta^{i_{2n+1} i_{2n+2}} \mu_{[n]}^{i_{2n+3}  \cdots i_s)} \hspace{.5cm}
\end{eqnarray} 
One finds that the coefficients $a_n$ are explicitly given by
\be
a_n = \frac{(-1)^n}{4^n} \frac{s}{n!} \frac{(s-n-1)!}{(s-2n)!}, \; \; \; (n\geq 1)
\ee
and that the coefficients $b_n$ are then
\be
b_n = a_n \frac{(s-2n) (s-2n-1)}{s(s-1)}, \; \; \;  (n \geq 1)
\ee
In index-free notations, the transformation of the Schouten tensor reads
\be 
\delta S = - d_{(s)}^2 \nu.
\ee

The recursive procedure amounts to successively eliminating the terms $\triangle \mu^{i_1 \cdots i_{s-2}}$, $\triangle \bar{\mu}^{i_1 \cdots i_{s-4}}$ involving the Laplacian by adding symmetrized products of $\delta^{ij}$'s with multiple traces of the Einstein tensors, with suitable coefficients that are determined uniquely. 

In terms of the variables $S^{i_1 i_2 \cdots i_s}$ and $\nu^{i_1 i_2 \cdots i_{s-2}}$, the Bianchi identity and the condition $\partial_{i_1} \mu^{i_1 i_2 \cdots i_{s-2}} = 0$ read respectively
\be
\partial_{i_1} S^{i_1 i_2 \cdots i_s} - (s-1) \partial^{(i_2}\bar{S}^{i_3 \cdots i_s)} =0 \label{BianchiForSs}
\ee
and
\be 
\partial_{i_1} \nu^{i_1 i_2 \cdots i_{s-2}} + \frac{s-3}{3} \partial^{(i_2} \bar{\nu}^{i_3 \cdots i_{s-2})} =0 \label{BianchiForNus}
\ee
The easiest way to prove these important relations is to observe that they follow uniquely from the requirement of invariance under (\ref{TransSchouten}), in much the same way as the Bianchi identity $\partial_{i_1} G^{i_1 i_2 \cdots i_s} = 0$ and the condition $\partial_{i_1} \mu^{i_1 i_2 \cdots i_{s-2}} = 0$ are the unique identity and condition compatible with the transformation (\ref{WeylTransG}) within the class
$$ \partial_{i_1} G^{i_1 i_2 \cdots i_s} +a \partial^{(i_2}\bar{G}^{i_3 \cdots i_s)}  + b \delta^{(i_2 i_3} \partial^{i_4} \bar{\bar{G}}^{i_5 \cdots i_s)} + \cdots =0,$$ 
$$ \partial_{i_1} \nu^{i_1 i_2 \cdots i_{s-2}} +k \partial^{(i_2}\bar{\nu}^{i_3 \cdots i_{s-2})}  + \ell \delta^{(i_2 i_3} \partial^{i_4} \bar{\bar{\nu}}^{i_5 \cdots i_{s-2})} + \cdots =0$$ 
(invariance under (\ref{WeylTransG}) forces $a=b=\cdots = k = \ell = \cdots = 0$).

\subsection{Cotton tensor}
In its original formulation, the Cotton tensor $C$ is defined as $d^{s-1}_{(s)} S$.  It is a tensor of mixed symmetry type
$$
C <> \underbrace{\overbrace{\yng(10,9)}^{\text{$s$ boxes}}}_{\text{$s-1$ boxes}} 
$$
which is invariant under spin-$s$ Weyl transformations as it follows from $d_{(s)}^{s+1} = 0$,
\be
\delta C = d_{(s)}^{s-1} \delta S = -d_{(s)}^{s+1} \nu = 0.
\ee
It contains $2s-1$ derivatives of the spin-$s$-field $h_{i_1 \cdots i_s}$.  As a consequence of the Bianchi identity,  it can be verified to be  traceless on the last index of the first row with any other index (i.e., one gets zero when the last index of the first row is contracted with any other index). 

In the dual representation on the first $s-1$ indices of $C$ which we shall adopt, the Cotton tensor $B^{i_1 \cdots i_s}$   is explicitly given by
\begin{eqnarray}
B^{i_1 i_2  \cdots i_s} &=& \varepsilon^{i_1 j_1 k_1}\varepsilon^{i_2 j_2 k_2} \cdots \varepsilon^{i_{s-1} j_{s-1} k_{s-1}} \nonumber \\
&& \hspace{.5cm} \partial_{j_1} \partial_{j_2} \cdots \partial_{j_{s-1}} S^{\hspace{1.38cm} i_s}_{k_1 k_2 \cdots k_{s-1}}
\end{eqnarray}
This tensor is manifestly symmetric in its first $s-1$ indices.  Symmetry in $i_{s-1}$, $i_s$ is a direct consequence of the Bianchi identity (\ref{BianchiForSs}) (this is equivalent to the tracelessness property of $C$ just mentioned).  Hence, the tensor $B^{i_1 \cdots i_s}$ is fully symmetric  i.e.,   is of symmetry type
$$
B <> \overbrace{\yng(10)}^{\text{$s$ boxes}} 
$$
Furthermore, it is easily proved to be conserved on the first index (i.e., its divergence on the first index is zero).  It is also traceless on the last two indices because of the Young symmetries of $C$.  Since $B$ is fully symmetric, one thus gets
\be
\delta_{i_k i_m} B^{i_1 i_2  \cdots i_s} = 0, \; \; \; \partial_{i_p } B^{i_1 i_2  \cdots i_s} = 0, 
\ee
 with $ 1 \leq  k<m\leq s $ and $1 \leq p \leq s$.
 
We stress that, as we have shown,  the Cotton tensor $B$ is completely symmetric as a consequence of the Bianchi identity.  Hence, it is not necessary to enforce symmetrization in its definition since it is automatic. Enforcing complete symmetrization, as done in \cite{Bergshoeff:2009tb,Bergshoeff:2011pm}, is of course permissible, but is not needed. 

\subsection{Spin-$2$}
\label{Sec:Spin2}
The above construction reproduces the familiar  spin-$2$ formulas.  One finds for the Schouten tensor,
$$ S^{ij} = G^{ij} - \frac12 \delta^{ij} \bar{G}, \; \; \;  G^{ij} = S^{ij} - \delta^{ij} \bar{S} $$
and $\delta S_{ij} = - \partial_i \partial_j \lambda$, $\partial_i S^{ij} - \partial^j \bar{S} = 0$.
The Cotton tensor $C_{ijk} $ is $C_{ijk} = \partial_i S_{jk} - \partial_j S_{ik} $
and is a Weyl-invariant tensor of type $\yng(2,1)$, which is traceless on $(j,k)$ (or $(i,k)$) because of the Bianchi identity, $\bar{C}_i = 0$.  It involves three derivatives of $h_{ij}$. In the dual representation, the Cotton tensor $B^{ij}$ is 
$$
B^{ij} =  \varepsilon^{i}_{\; \, mn} \partial^{m} S^{nj} 
$$
and is easily checked to be indeed symmetric, traceless and divergenceless.

\subsection{Spin-$3$}

We now move to the spin 3 case,
$$
h <>\yng(3) ,
$$
where the above derivation reproduces  the results of \cite{Damour:1987vm}. We derive below the form the general formulas take for $s=3$.

The Schouten tensor for a spin-$3$ field reads explicitly
\be
S^{i_1 i_2 i_3} = G^{i_1 i_2 i_3} - \frac{3}{4} \delta^{(i_1 i_2} \bar{G}^{i_3)}  \label{Schouten3}
\ee
Its trace is equal to
$
\bar{S}^i = -\frac14 \bar{G}^i
$ so that the inverse formula to (\ref{Schouten3}) is
$
G^{i_1 i_2 i_3} = S^{i_1 i_2 i_3} - 3 \delta^{(i_1 i_2} \bar{S}^{i_3)} 
$.
The Schouten tensor transforms as
\be
\delta S_{i_1 i_2 i_3} = -3 \partial_{(i_1} \partial_{i_2} \mu_{i_3)}
\ee
under Weyl transformations, where $\mu_i$ is given by
\be
\mu^i =\varepsilon^{ijk} \partial_j \lambda_k
\ee
and fulfills $\partial_k \mu^k = 0$.

The Bianchi identity implies $\partial_i \bar{S}^i = 0$ and can  equivalently be written in terms of $S^{i_1 i_2 i_3}$ as
\be
\partial_{i_1} S^{i_1 i_2 i_3} - \partial^{i_2} \bar{S}^{i_3} - \partial^{i_3} \bar{S}^{i_2} = 0 \label{BianchiSSpin3}
\ee

According to the above general definition, the Cotton tensor $C= d_{(3)}^2 S$ is explicitly given by 
\begin{eqnarray}
&& C_{i_1 j_1 \vert i_2 j_2 \vert i_3} = \partial_{i_1} \partial_{i_2} S_{j_1 j_2 i_3} - \partial_{j_1} \partial_{i_2} S_{i_1 j_2 i_3} \nonumber \\ 
&& \hspace{1cm} - \partial_{i_1} \partial_{j_2} S_{j_1 i_2 i_3} + \partial_{j_1} \partial_{j_2} S_{i_1 i_2 i_3} \nonumber
\end{eqnarray}
and has Young symmetry type
$$
C<>\yng(3,2) 
$$
In the dual representation, the Cotton tensor $B^{k_1 k_2 k_3}$ reads
\be 
B^{k_1 k_2 k_3} = \varepsilon^{k_1 i_1 j_1} \varepsilon^{k_2 i_2 j_2}  \partial_{i_1} \partial_{i_2} S_{j_1 j_2}^{\; \; \; \; \; \; k_3} \label{DualCottonSpin3}
\ee
and, using the Bianchi identity (\ref{BianchiSSpin3}) and its consequence $\partial_j \bar{S}^j = 0$, is easily seen to be equal to 
\be
B^{k_1 k_2 k_3} = 3 \partial^{(i_1} \partial^{i_2} \bar{S}^{i_3)} - \triangle S^{i_1 i_2 i_3}
\ee
an expression that is manifestly symmetric.
Transverseness and tracelessness follow again from the Bianchi identity (\ref{BianchiSSpin3}).

\subsection{Spin-$4$}

We now write the formulas in the spin-$4$ case. The
Schouten tensor is
\be
S^{i_1 i_2 i_3 i_4} = G^{i_1 i_2 i_3 i_4} -  \delta^{(i_1 i_2} \bar{G}^{i_3  i_4)} + \frac{1}{8}  \delta^{(i_1 i_2} \delta^{i_3 i_4)} \bar{\bar G} 
\ee
and transforms as
\be
\delta S_{i_1 i_2 i_3 i_4} = - 6 \partial_{(i_1} \partial_{i_2} \nu_{i_3 i_4)}
\ee
under Weyl transformations, with $\nu_{ij} = \mu_{ij} - \frac16 \delta_{ij} \bar{\mu}$.

In terms of the Schouten tensor, the Bianchi identity reads
\be
\partial_{i_1} S^{i_1 i_2 i_3 i_4} - \partial^{i_2} \bar{S}^{i_3 i_4} - \partial^{i_3} \bar{S}^{i_2 i_4} -  \partial^{i_4} \bar{S}^{i_2 i_3} = 0 \label{BianchiSpin4a}
\ee
and one has $\partial_{i} \nu^{ij }+ \frac13 \partial^j \bar{\nu} = 0$.

The spin-$4$ Weyl invariant Cotton tensor is $d^3_{(4)} S$. Writing the explicit formulas directly in the dual representation, one finds
\be 
B^{k_1 k_2 k_3 k_4} = \varepsilon^{k_1 i_1 j_1} \varepsilon^{k_2 i_2 j_2}  \varepsilon^{k_3 i_3 j_3} \partial_{i_1} \partial_{i_2} \partial_{i_3} S_{j_1 j_2 j_3}^{\; \; \; \; \; \; \; \; k_4} \, . \label{DualCottonSpin4}
\ee
 Again, the symmetry in $(k_1, k_2, k_3)$ is manifest, while the symmetry in the last index $k_4$ with any other index is a consequence of the Bianchi identity (\ref{BianchiSpin4a}).  The Cotton tensor is transverse and traceless,
\be
\partial_i B^{ijkl} = 0, \; \; \; \bar{B}^{ij} = 0.
\ee

\section{Higher spin ``Conformal flatness"}
\setcounter{equation}{0}
\label{Sec:4}

The Cotton tensor is quite important because it completely captures higher spin Weyl invariance.  By this, we mean that any function of the higher spin field and its derivatives that is invariant under higher spin diffeomorphisms and Weyl transformations is necessarily a function of the Cotton tensor and its derivatives,
\be
\delta_{\xi, \lambda}  f([h]) = 0 \; \; \Rightarrow f= f([B])
\ee
Equivalently, a necessary and sufficient condition for a spin-$s$ field to be pure gauge (equal to zero up to spin-$s$ diffeomorphisms and Weyl transformations) is that its Cotton tensor vanishes.

The first version of this property is demonstrated in Appendix {\bf \ref{App:Complete}}.  We show here how to prove the second version\footnote{We were kindly informed by Xavier Bekaert that the property ``Cotton tensor = 0 $\Leftrightarrow$ spin-$s$ field is diffeomorphism and Weyl pure gauge" can also be viewed as a consequence of the cohomological theorems of \cite{Shaynkman:2004vu} on the representations of the conformal group, see \cite{Beccaria:2014jxa,BekaertChengdu}. We are grateful to him for this information.}.  

Assume, then,  that the Cotton tensor $B$ (or equivalently, $C$) is equal to zero.  Using the cohomological theorems of \cite{DuboisViolette:1999rd}, one gets 
\be
C = d_{(s)}^{s-1} S = 0  \Rightarrow S = - d_{(s)}^2 \nu
\ee for some $\nu$. The Bianchi identity implies that one can choose $\nu$ in such a way that the corresponding tensor $\mu$ fulfills $\partial_{\mu_1} \mu^{i_3 \cdots i_s} = 0$ and so,  can be written as in (\ref{FormOfMu}) for some completely symmetric $\lambda_{k_3  \cdots k_s}$.
This implies in turn that $R[h-  \lambda \star \delta] = 0$, or equivalently $d_{(s)}^s (h - \lambda \star \delta) = 0$.  Here, $ \lambda \star \delta$ stands for the Weyl transformation term $\delta_{(i_1 i_2} \lambda_{i_3 \cdots i_s)}$.  Using again the cohomological theorems of \cite{DuboisViolette:1999rd}, one finally obtains
\be
h = d_{(s)}  \xi +  \lambda \star \delta,
\ee
i.e., $h$ is pure gauge.  Conversely, if $h$ is pure gauge, the Cotton tensor vanishes.
We can thus conclude that a necessary and sufficient condition for the spin-$s$ field to be pure gauge is that its Cotton tensor vanishes.

We illustrate explicitly the derivations in the spin-3 and spin-4 cases.

\subsection{Spin $3$}

Consider a spin-$3$ field $h_{i_1 i_2 i_3}$ with vanishing  Cotton tensor.  According to the theorems of \cite{DuboisViolette:1999rd}, the Schouten tensor reads
\be
 S_{i_1 i_2 i_3} = - \partial_{(i_1} \partial_{i_2} \mu_{i_3)} \label{EqForS}
\ee
for some $\mu_i$.  We want to prove that $\mu_i$ can be chosen so that $\partial_i \mu^i = 0$.  From the Bianchi identity, one gets 
$$
\partial_i \partial_j (\partial_k \mu^k) = 0
$$
It follows that $\partial_k \mu^k$ is at most linear in the coordinates,
$$
\partial_k \mu^k = a + b_k x^k
$$
Define $\tilde{\mu}^k = \frac13 a x^k + \frac 12 c^k_{\; \; ij} x^i x^j$ where $ c^k_{\; \; ij} =  c^k_{\; \; ji}$, $ c_{(kij)} = 0$ and $ c^k_{\; \; kj} = b_j$. [Such a $ c^k_{\; \; ij}$ exists.  It has the Young symmetry $\yng(2,1)$ in the dual conventions where symmetry is manifest while antisymmetry is not.  The trace of such a tensor is unconstrained and so can be taken to be equal to $b_j$.]  By construction, $\partial_k \tilde{\mu}^k = \partial_k \mu^k$ and $\partial_{(i_1} \partial_{i_2} \tilde{\mu}_{i_3)}=0$, so that $S_{i_1 i_2 i_3} = - \partial_{(i_1} \partial_{i_2} (\mu_{i_3)} - \tilde{\mu}_{i_3)})$, implying that we can assume that $\mu^k$ in (\ref{EqForS}) fulfills $\partial_k \mu^k = 0$, which will be done from now on.  We then have $\mu^k = \varepsilon^{kij} \partial_i \lambda_j$ for some $\lambda_j$, and so the Einstein tensor of $h_{ijk}$ is equal to the Einstein tensor of $3 \delta_{(ij} \lambda_{k)}$, implying $h_{ijk} = 3 \partial_{(i} \xi_{jk)} + 3 \delta_{(ij} \lambda_{k)}$ for some $\xi_i$, as announced in the general discussion above.

\subsection{Spin $4$}

We now illustrate the produre for the spin-$4$ field.  The vanishing of the Cotton tensor implies again 
\cite{DuboisViolette:1999rd}, 
\be
 S_{i_1 i_2 i_3 i_4} = - \partial_{(i_1} \partial_{i_2} \nu_{i_3 i_4)} \label{EqForS4}
\ee
for some $ \nu_{i_3 i_4}$.  The Bianchi identity yields then
\be
\partial_{(i}\partial_j N_{k)} = 0, \; \; \; N_{k} \equiv \partial^m \nu_{km} + \frac13 \partial_k \bar{\nu}.
\ee
This does not imply that $N_k =0$ since, as for spin-$3$, there are non trivial solutions of the equation $\partial_{(i}\partial_j N_{k)} = 0$.   These solutions have been analyzed in section {\bf 6} of \cite{DuboisViolette:2001jk}.   The space of solutions is finite-dimensional; one easily gets from the equation that the third derivatives of $N_k$ vanish, so that $N_k$ is at most quadratic in the $x^i$'s,
$$ N_k = a_k + b_{k\vert m} x^m + c_{k \vert mn}x^m x^n,$$
for some constants $a_k$,  $b_{k\vert m}$ and $ c_{k \vert mn} = c_{k \vert nm}$ which have the respective symmetry $\yng(1)$, $\yng(1) \otimes \yng(1)$, and $\yng(1) \otimes \yng (2)$.  Now, let 
$$
\tilde{\nu}_{km} = \rho (a_k x_m + a_m x_k) + \sigma_{km \vert rs} x^r x^s  + \theta_{km \vert rsp} x^r x^s x^p
$$
where the constants $\rho$, $\sigma_{km \vert rs}$ (with symmetry $\yng(2) \otimes \yng(2)$) and $\theta_{km \vert rsp}$ (with symmetry $\yng(2) \otimes \yng(3)$) are chosen such that (i) $\sigma_{(km \vert ij)}=0$, $\theta_{(km\vert ij)p} = 0$ so that $\partial_{(i}\partial_j \tilde{\nu}_{km)} = 0$; and (ii) $\tilde{N}_k = N_k$.  This is always possible since this second condition restricts only the traces, which are left free by the first condition.  Then, by substracting $\tilde{\nu}_{km}$ from $\nu_{km}$, one sees that one can assume $N_k=0$.  This implies that the corresponding $\mu^{km}$ can be assumed to fulfill $\partial_m \mu^{km} = 0$ and thus is equal to $\mu^{km} = \varepsilon^{krs} \varepsilon^{mpq} \partial_r \partial_p \lambda_{sq}$ for some $\lambda_{sq} = \lambda_{qs}$.  Therefore, the Einstein tensor  of $h_{ijkm}$ is equal to the Einstein tensor of $\delta_{(ij} \lambda_{km)}$, implying that
$$h_{ijkm} = 4\partial_{(i} \xi_{jkm)} + 6 \delta_{(ij} \lambda_{km)},$$
which is the result that we wanted to prove.

\section{A crucial property}
\setcounter{equation}{0}
\label{Sec:Crucial}

We have proven so far the analogs of properties 1-5 for the higher spin Cotton tensors.  We turn now to property 6.

\subsection{The problem}
We have recalled that if a completely symmetric tensor $G^{i_1 \cdots i_s}$ fulfills the equation
\be
\partial_{i_1} G^{i_1 i_2 \cdots i_s} = 0,
\ee
then there exists $h_{i_1 \cdots i_s}$ such that $G = G[h]$.

We want to address the question: let $B^{i_1 i_2 \cdots i_s}$ be a completely symmetric tensor that is both transverse 
\be
\partial_{i_1} B^{i_1 i_2 \cdots i_s} = 0,  \label{EQ1}
\ee
and traceless,
\be
\delta_{i_1 i_2} B^{i_1 i_2 i_3\cdots i_s} = 0, \label{EQ2}
\ee
Does there exist a totally symmetric tensor $Z_{i_1 \cdots i_s}$ such that $B^{i_1 i_2 \cdots i_s}$ is the Cotton tensor of $Z_{i_1 \cdots i_s}$?

We prove here that  the answer is affirmative, starting with the spin-$2$ case.

\subsection{Spin $2$}

The Cotton tensor $B^{ij}$ is dual to $C_{ijk} = - C_{jik}$ on the first index.  The tracelessness condition on $B^{ij}$ implies that $C_{ijk}$ has Young symmetry type $\yng(2,1)$, while the symmetry in $(i,j)$ of $B^{ij}$ implies that $C_{ijk}$ is traceless. The divergenceless condition on $B^{ij}$ implies then, by Poincar\'e lemma ($k$ being a ``spectator" index) that
\be
C_{ijk} = \partial_i S_{jk} - \partial_j S_{ik}
\ee
where $S_{ik}$ is not a priori symmetric.  However, the ambiguity in $S_{ik}$ is $S_{ik} \rightarrow S_{ik} + \partial_i T_k$,  and using this ambiguity, the condition $C_{[ijk]} = 0$ and Poincar\'e lemma, one easily sees that $S_{ik}$ can be assumed to be symmetric.  Then, the tracelessness condition implies the Bianchi identity $\partial_i S^{ij} - \partial^j \bar{S} = 0$ for $S^{ij}$ (or $\partial_i G^{ij} = 0$ for $G^{ij}$), from which follows the existence of $Z_{ij}$ such that $S = S[Z]$ and thus $B = B[Z]$.  This establishes the result.

\subsection{Higher spin}
The same steps work for higher spins.  For instance, for spin $3$, the reasoning proceeds as follows:
\begin{itemize}
\item Define $C_{i_1 j_1 \vert i_2 j_2 \vert k}$ from $B^{ijk}$ by dualizing on the first two indices, with $k$ ``spectator",
\be
C_{i_1 j_1 \vert i_2 j_2 \vert k}= \epsilon_{i_1 i_2 i} \epsilon_{j_1 j_2 j} \delta_{km} B^{ijm}
\ee
\item Because $B^{ijm}$ is completely symmetric and traceless, $C_{i_1 i_2 \vert j_1 j_2 \vert k}$ has Young symmetry type $\yng(3,2)$ and is traceless on $k$ and any other of its indices.
\item The transverse condition on $B^{ijk}$ is equivalent to  $d_{(2)} C = 0$ where $d_{(2)}$ is acting on $C$ as if it was a collection of tensors of type $\yng(2,2)$ parametrized by $k$. The Poincar\'e lemma implies then the existence of a tensor $R_{ijk}$ such that 
\be C_{i_1 j_1 \vert i_2 j_2 \vert k}=  \partial_{[i_1} \partial_{[j_1} R_{i_2] j_2] k} \label{FormOfC}
\ee
where the antisymmetrizations 	are on the pairs of indices $(i_1, i_2)$ and $(j_1,j_2)$. At this stage, the tensor $R$ has a component $S$ that is completely symmetric $S <> \yng(3)$,
\begin{eqnarray}
S_{i_2 j_2 k} &=& \frac13(R_{i_2 j_2 k}  + R_{ j_2 k i_2}  + R_{k i_2 j_2} ) \\
&=& S_{(i_2 j_2 k)} 
\end{eqnarray}
and a component $T$ that has Young symmetry type $\yng(2,1)$,
\be
T_{ijk} = \tilde{T}_{kji} - \tilde{T}_{kij}
\ee
\be
3\tilde{T}_{kji} = T_{ijk} +T_{ikj}
\ee
with
\be
\tilde{T}_{i_2 j_2 k} = 2 R_{i_2 j_2 k} - R_{k j_2 i_2} - R_{k i_2 j_2}
\ee
Explicitly,
\be
R_{i_2 j_2 k} = S_{i_2 j_2 k} + \frac13 \tilde{T}_{i_2 j_2 k}.
\ee
The change from $T$ to $\tilde{T}$ corresponds to the change of conventions in which either antisymmetry or symmetry is manifest.  We shall call $T$ the ``Curtright tensor".
\item The tensor $R_{i_2 j_2 k}$ is not completely determined by $C_{i_1 j_1 \vert i_2 j_2 \vert k}$ since one may add to it 
\be
R_{i_2 j_2 k} \rightarrow R_{i_2 j_2 k} + \partial_{i_2} \mu_{j_2 k} + \partial_{j_2} \mu_{i_2 k} \label{GaugeR}
\ee
without violating (\ref{FormOfC}). This is the only ambiguity.
\item Furthermore, the condition $C_{i_1 j_1 \vert [i_2 j_2 \vert k]} = 0$ is easily verified to imply that the field strength of the Curtright tensor is equal to zero, so that $T$ is pure gauge and can be set equal to zero by a gauge transformation of the type (\ref{GaugeR}).  
Therefore, one can assume
\be C_{i_1 j_1 \vert i_2 j_2 \vert k}=  \partial_{[i_1} \partial_{[j_1} S_{i_2] j_2] k} \label{FormOfCbis}
\ee
with $S$ completely symmetric.
\item The residual gauge symmetry after $T$ has been set equal to zero is given by
\be
S_{ijk} \rightarrow S_{ijk} + \partial_{(i} \partial_j \mu_{k)}  \label{Residual}
\ee (which is still present because the gauge symmetries of the Curtright tensor are reducible).
\item Finally, the tracelessness condition of $C$ on $k$ and any other index yields
\be
\partial_{[i_1} U_{i_2]j} = 0 \label{BianchiU}
\ee
where 
\be
U_{ij} = \partial^k S_{ijk} - \partial_i \bar{S}_j - \partial_j \bar{S}_i
\ee
is such that $U_{ij} =0$ is the Bianchi identity for the Schouten tensor (see (\ref{BianchiSSpin3})).   Now, (\ref{BianchiU}) implies
\be
U_{ij} = \partial_{i} \partial_{j} \rho
\ee 
for some $\rho$ ( cohomology of $d_{(2)}$ for $d_{(2)}$ with $d_{(2)}^3 = 0$).   On the other hand, $U_{ij}$ transforms as
\be
U_{ij} \rightarrow U_{ij} - 3 \partial_i \partial_{j} (\partial_k \mu^k)
\ee
under (\ref{Residual}).  This enables one to chose $S$ to obey  the Bianchi identity of the Schouten tensor (take $\mu^k$ such that $3\partial_k \mu^k = \rho$), implying the existence of $Z_{ijk}$ such that $S = S[Z]$ and hence $B = B[Z]$. [Note that $\delta U =0$ when $\partial_k \mu^k =0$, as it should.]  This ends the demonstration of  the property that we wanted to prove.
\end{itemize}

\section{Prepotentials and solution of the constraints of the Hamiltonian formulation}
\setcounter{equation}{0}
\label{Sec:Prepo}

The dynamics of bosonic higher spin gauge fields in four spacetime dimensions is given by the Fronsdal action \cite{Fronsdal:1978rb} expressed in terms of a completely symmetric spacetime field 
 $h_{\mu_1 \cdots \mu_s}$ which is subject to the double trace condition.  The gauge symmetries read
\be
\delta  h_{\mu_1 \cdots \mu_s} = s\partial_{(\mu_1} \epsilon_{\mu_2 \cdots \mu_s)}
\ee
where the gauge parameter $\epsilon_{\mu_2 \cdots \mu_s}$ is subject to the single trace condition.  

Because of the single trace condition, the gauge parameters are not independent.  One can take $\epsilon_{i_1 \cdots i_{s-1}}$ and $\epsilon_{0 i_1 \cdots i_{s-2}}$ as independent gauge parameters, since two subscripts $0$ in the gauge parameters can be replaced by spatial indices through the trace condition.

Similarly, one can express through the double trace conditions all components of the spin-$s$ field with 4 or more subscripts $0$ in terms of $h_{i_1 \cdots i_s}$, $h_{0 i_2 \cdots i_{s-1}}$, $h_{00i_3 \cdots i_{s-2}}$ and $h_{000i_1 \cdots i_{s-3}}$.

In the transition to the Hamiltonian formalism worked out in \cite{Metsaev:2011iz} (see also \cite{CHHL} for a discussion that includes the analysis of the surface terms), the variables $h_{0 i_2 \cdots i_{s-1}}$ and $h_{00i_3 \cdots i_{s-2}}$ play the role of Lagrange multipliers for the constraints associated with the independent gauge parameters $\epsilon_{i_1 \cdots i_{s-1}}$ and $\epsilon_{0 i_1 \cdots i_{s-2}}$, while $h_{i_1 \cdots i_s}$ and $\alpha_{i_1 \cdots i_{s-3}} \equiv h_{000i_1 \cdots i_{s-3}} - 3 h^k_{\; k0i_1 \cdots i_{s-3}}$, together with their conjugate momenta $\pi^{i_1 \cdots i_s}$ and $\Pi^{i_1 \cdots i_{s-3}} $ are the (constrained) phase space variables.  The constraints are of second order in the variables and of first order in their momenta and split into two groups, the ``Hamiltonian constraints" ${\mathcal H}_{i_1 \cdots i_{s-2}} \approx 0 $ associated with the gauge parameters $\epsilon_{0 i_1 \cdots i_{s-2}}$
and the ``momentum constraints" ${\mathcal H}_{i_1 \cdots i_{s-1}} \approx 0$ associated with the gauge parameters $\epsilon_{i_1 \cdots i_{s-1}}$.
The Hamiltonian is quadratic in the conjugate momenta and in the derivatives of the fields.  

The explicit expressions for the spin-$3$ case, with phase space variables $h_{ijk}$, $\alpha$, $\pi^{ijk}$, $\Pi$, are respectively
\begin{eqnarray}
H &=& \int d^3 x \left\lbrace
\frac{1}{2} \Pi_{ijk} \Pi^{ijk}
 - \frac{3}{8} \bar{\Pi}_k \Pi^k
 + \frac{3}{8} \bar{\Pi}^k \partial_k \alpha
 + \frac{17}{32} \partial_k \alpha \partial^k \alpha + \Pi^2 \right. \nonumber\\ \qquad 
&&\left. 
 + \frac{1}{2} \partial_{k} h_{lmn}\partial^{k} h^{lmn}
 - \frac{3}{2}  \partial_{k} h_{lmn} \partial^{l} h^{kmn}
 + 3 \partial^{l} h_{klm} \partial^{k} \bar{h}^{m}
 - \frac{3}{2} \partial_{k} \bar{h}_{l} \partial^{k} \bar{h}^{l}
 - \frac{3}{4} \partial_{k} \bar{h}_{l} \partial^{l} \bar{h}^{k} \right\rbrace\nonumber\\ \label{HamiltonianSpin3}
\end{eqnarray}
(Hamiltonian),
\be
{\mathcal H}_{i} = \partial_i \Pi - \triangle \bar{h}_i  + \partial^j \partial^k h_{ijk}  - \frac12 \partial_i \partial^j \bar{h}_j \approx 0
\ee
(Hamiltonian constraint) and
\be
{\mathcal H}_{ij} = 2 \partial^k \pi _{ijk} + \delta_{ij} \triangle \alpha \approx 0
\ee
(momentum constraint).

The Hamiltonian constraint generates the gauge transformations
\begin{eqnarray}
\hspace{-.4cm} &&\delta_\chi \pi^{ijk} = -  \partial^{(i} \partial^j \chi^{k)} +  \delta^{(ij}  \left(  \triangle \chi^{k)} + \frac12 \partial^{k)} \partial_m \chi^m \right)\hspace{.7cm}\\
\hspace{-.4cm}&& \delta_\chi \alpha = -  \partial_m \chi^m  
\end{eqnarray}
($\delta_\chi h_{ijk} = 0$, $\delta_\chi \Pi = 0$)
which are the ``temporal spin-$3$ diffeomorphisms" with gauge parameters $\chi_i \sim \epsilon_{0 i}$, while the momentum constraint generates spatial spin-$3$ diffeomorphisms with gauge parameters $\epsilon_{ij}$,
\begin{eqnarray}
&& \delta_\epsilon h_{ijk} = 3 \partial_{(i} \epsilon_{jk)}, \\
&& \delta_\epsilon \Pi = \frac32 \triangle \bar{\epsilon}
\end{eqnarray}
($\delta_\epsilon \pi^{ijk} = 0$, $\delta_\epsilon \alpha = 0$).  The use of the terminology ``Hamiltonian constraint" and ``momentum constraint" is motivated by the spin-$2$ case.

\subsection{Momentum constraint}
We first solve the momentum constraint. Using the $\chi$-gauge transformations, one can set $\alpha = 0$. In that gauge, the constraint reduces to $ \partial_i \pi^{ijk} = 0$, which implies $\pi^{ijk} = G^{ijk}[P]$ for some prepotential $P_{ijk}$, which is at this stage determined up to a spin-$3$ diffeomorphism.

In a general gauge, one has therefore
\begin{eqnarray}
&&\pi^{ijk} = G^{ijk}[P]  -  \partial^{(i} \partial^j \Xi^{k)} \nonumber \\
&& \hspace{1cm} +  \delta^{(ij}  \left(  \triangle \Xi^{k)} + \frac12 \partial^{k)} \partial_m \Xi^m \right)  \label{PrepoForpi}\\
&&\alpha = -  \partial_m \Xi^m 
\end{eqnarray}
where $\Xi_k$ is a second prepotential that describes the gauge freedom of $\pi^{ijk}$ and $\alpha$.

Now, the vector $\Xi^k$ can be decomposed into a transverse and a longitudinal piece,
$$\Xi^k = \varepsilon^{kij} \partial_i \lambda_j + \partial^k \theta.$$  The $\lambda^k$-terms in (\ref{PrepoForpi}) are easily checked to be of the form $G^{ijk}[\varphi]$, where $\varphi_{ijk}=\delta_{(ij} \lambda_{k)}$ has just the form of a spin-$3$ Weyl transformation.  This shows that the prepotential $P_{ijk}$ is determined up to a spin-$3$ Weyl transformation -- in addition to the spin-$3$ diffeomorphism invariance pointed out above.  Therefore, the gauge freedom of the prepotential is
\be
\delta P_{ijk} = 3 \partial_{(i} \xi_{jk)} + 3 \delta_{(ij} \lambda_{k)},
\ee
i.e., the gauge symmetries of a conformal spin-$3$ field.

The fact that the spin-$3$ diffeomorphisms of the prepotential $P_{ijk}$ have no action on the canonical variables, while its conformal transformations generate (some of) the gauge transformations associated with the Hamiltonian constraint, parallels the situation found in the case of spin 2 \cite{Henneaux:2004jw,Bunster:2012km,Bunster:2013oaa}.  There, however, the Weyl transformations of the prepotential accounted for {\em all} the gauge symmetries generated by the Hamiltonian constraint. 

\subsection{Hamiltonian constraint}

We now turn to solving the Hamiltonian constraint.    Its curl $\epsilon^{ijk} \partial_j {\mathcal H}_k $ does not involve $\Pi$ and turns out to be equal to $\bar{G}^i$, so that the Hamiltonian constraint implies
\be 
\bar{G}^i[h] = 0.
\ee
In fact, one may rewrite the Hamiltonian constraint as
\be
\partial_i \Pi - \Psi_i = 0
\ee 
in terms of the $\Psi$ introduced in SubSection {\bf \ref{SubSec:Einstein}}, such that $^* \bar{G} = d \Psi$.  One has explicitly
\be
\Psi_i = \triangle \bar{h}_i - \partial^j \partial^k h_{ijk} + \frac12 \partial_i \partial^j \bar{h}_j
\ee
Therefore, the equations $\bar{G}^i =0 \Leftrightarrow d \Psi = 0$ and $d \Pi + \Psi=0$ are two equivalent versions of the Hamiltonian constraint.  

The form $\bar{G}^i = 0$  is more amenable to solution because it falls precisely under the analysis of Section {\bf \ref{Sec:Crucial}}.  According to what we have proved there, it implies the existence of a (second) prepotential $\Phi_{ijk}$ such that the Einstein tensor of $h$ is the Cotton tensor of that prepotential,
\be G^{ijk}[h] = B^{ijk}[\Phi] \label{G=B}
\ee
A particular solution of (\ref{G=B}) is given by
\begin{eqnarray}
h_{ijk} &=& - \triangle \Phi_{ijk} + \frac34 \delta_{(ij} \triangle \bar{\Phi}_{k)} \nonumber \\
&& - \frac34 \delta_{(ij} \partial^r \partial^s \Phi_{k)rs} + \frac{3}{10} \delta_{(ij} \partial_{k)} \partial^r \bar{\Phi}_r . \label{SolG=B}
\end{eqnarray}
The last term in (\ref{SolG=B}) is not necessary but included so that $\delta h_{ijk} = 0$ under Weyl transformation of $\Phi$. 

Now, what are the ambiguities? It is clear that the spin-$3$ field $h_{ijk}$ is determined by (\ref{G=B})  up to a spin-$3$ diffeomorphism, so that the general solution of (\ref{G=B}) is given by (\ref{SolG=B}) plus $\partial_{(i} u_{jk)}$ where $u_{jk}$ may be thought of as another prepotential that drops out because of gauge invariance.   Conversely, the prepotential $\Phi_{ijk}$  itself is determined by (\ref{G=B}), i.e., by its Cotton tensor, up to a diffeomorphism and a Weyl rescaling,
\be
\Phi_{ijk} \rightarrow \Phi_{ijk} + 3 \partial_{(i} \xi'_{jk)} + 3 \delta_{(ij} \lambda'_{k)} \label{GaugePhi}
\ee
with independent gauge parameters $\xi'_{ij}$ and $\lambda'_j$. Thus, we see that the resolution of the Hamiltonian constraints also introduces a prepotential possessing the gauge symmetries of a conformal spin-$3$ field.  Note that we have adjusted the ambiguity in the dependence of $h_{ijk}$ on $\Phi_{ijk}$ in such a way that the conformal spin-$3$ transformations of the prepotential leave $h_{ijk}$ invariant, while the spin-$3$ diffeomorphisms of the prepotential induce particular spin-$3$ diffeomorphisms of $h_{ijk}$, as is the case for spin 2 \cite{Henneaux:2004jw,Bunster:2012km,Bunster:2013oaa}.

Once $h_{ijk}$ is determined, one may work one's way up to the constraint and solve for $\Pi$ in terms of the prepotential.  One finds \be
\Pi = - \frac18 \partial^{i}\partial^j\partial^{k} \Phi_{ijk} + \frac{3}{40} \triangle \partial^i \bar{\Phi}_i \ee

The use of conformal techniques to solve the Hamiltonian constraint is somewhat reminiscent of the approach to the initial value problem for full general relativity developed in  \cite{Arnowitt:1962hi,Deser:1967zzb,York:1971hw}.

\section{Manifest duality invariance}
\setcounter{equation}{0}
\label{Sec:Duality}

If one rewrites the action in terms of the prepotentials $(Z^a_{ijk}) \equiv (P_{ijk}, \Phi_{ijk})$ ($a= 1,2)$, one finds the remarkable simple expression
\be
S = \int dx^0 \left[ \int d^3x \ \frac{1}{2}\varepsilon_{ab} B^{a\, ijk} \dot{Z}^b_{ijk} - H \right] \label{ActionPrepot00}
\ee
where the Hamitonian $H$ reads
\be
H = \int d^3x \delta_{ab} \left(\frac{1}{2}  G^{a \, ijk}G^b_{ijk} - \frac{3}{8} \bar{G}^{a \, i} \bar{G}^b_i \right)
\ee
 Here, $\varepsilon_{ab}$ and $\delta_{ab}$ are respectively the Levi-Civita tensor and the Euclidean metric in the internal plane of the two prepotentials, while $G^a_{ijk} \equiv G_{ijk}[Z^a]$ and $B^a_{ijk} \equiv B_{ijk}[Z^a]$.  In terms of the prepotentials, the action possesses exactly the same structure as the action for spin 2 \cite{Bunster:2012km}.

The kinetic term in the action is manifestly invariant under the gauge symmetries of the prepotentials. The Hamiltonian is  manifestly invariant under the spin-$3$ diffeomorphisms, since it involves the Einstein tensors of the prepotentials.  It is also invariant under spin-$3$ Weyl transformations up to a surface term, as it can easily be verified.

The action is furthermore manifestly invariant under $SO(2)$ electric-magnetic duality rotations in the internal plane of the prepotentials, 
\begin{eqnarray}
\Phi'^{ijk} &=& \cos \theta \Phi^{ijk} - \sin \theta P^{ijk} ,
\\
P'^{ijk} &=& \cos \theta P^{ijk} + \sin \theta\Phi^{ijk}
\end{eqnarray}
since it involves only the $SO(2)$ invariant tensors $\varepsilon_{ab}$ and $\delta_{ab}$.  As recalled in the introduction, exhibiting duality symmetry in the case of spin $2$ was in fact the main motivation of \cite{Henneaux:2004jw} for solving the constraints and introducing the prepotentials.

The gauge symmetries combined with duality invariance constrain the form of the action in a very powerful way. Indeed, the most general invariant quadratic kinetic term involving 6 derivatives of the prepotentials, among which one is a time derivative,  is a multiple of the above kinetic term.  Similarly, the most general invariant quadratic Hamiltonian involving 6 spatial derivatives of the prepotentials is  a mutiple of the above Hamiltonian.   By rescaling appropriately the time if necessary, one can therefore bring the action to the above form, which is consequently the most general gauge and duality invariant quadratic action with the required number of derivatives.

In terms of the prepotentials, the equations of motion read
\be
\dot{B}^a_{ijk} = \epsilon^{ab} \epsilon_{ilm} \partial^l G_{b\, jk}^{\ \ \ m}  \label{EOM3}
\ee
and equate the time derivative of the Cotton tensor of one prepotential to the ``curl" of the Cotton tensor of the other (defined as the right-hand side of (\ref{EOM3})).

\section{Comments and Conclusions}
\label{Conclusions}
\setcounter{equation}{0}

Perhaps the most intriguing feature of our analysis is the emergence of spin-$3$ Weyl gauge invariance.  Starting from the ordinary spin-$3$ Fronsdal Lagrangian, which  exhibits no sign of  higher spin conformal gauge symmetry, the resolution of the constraints of the Hamiltonian formalism brings in prepotentials that enjoy somewhat unexpectedly this symmetry.  This feature was already found in the spin-$2$, spin $3/2$ and spin-$5/2$ contexts \cite{Henneaux:2004jw,Bunster:2012jp,Bunster:2013oaa,Bunster:2014fca} and will be confirmed in even higher spin models, both bosonic \cite{HHL0} and fermionic \cite{HHL} -- papers in which we shall also discuss the twisted self-duality formulation of higher spins in terms of electric and magnetic fields.  This is what justifies the construction of the appropriate conformal calculus (Cotton tensor) in the present context.
A similar emergence of local higher spin conformal symmetry arises in higher dimensions, and the prepotentials appear to be systematically of a Young symmetry type such that the corresponding Weyl tensor identically vanishes so that one must go to the analog of the Cotton tensor \cite{Bunster:2013oaa,HHL0}.  

The ultimate reason for the emergence of local higher spin Weyl symmetry remains to be understood.  This seems to us to be particularly important in view of the power of this symmetry which determines, together with higher spin diffeomorphisms and $SO(2)$ duality invariance, the form of the action.   In that respect, it should be noted that although not manifestly so, the action compatible with all the listed symmetries is automatically also Lorentz invariant since it is equivalent to the Fronsdal action.  This is in line with \cite{Bunster:2012hm} (see also \cite{Gelfond:2015poa} in this context).

The simplicity of the action (\ref{ActionPrepot00}) should be contrasted with its expression in terms of the original variables.  In particular, the Hamiltonian expressed in terms of the prepotentials is much more transparent than its original expression (\ref{HamiltonianSpin3}). A similar simplicity also holds for spins $>3$, where the action is found to have the same universal form, with a kinetic term involving the time derivative of the Cotton tensor, and a Hamiltonian quadratic in the Riemann tensor and its multiple traces, with coefficients that ensure higher spin conformal symmetry.

Finally, one can trade off the prepotential $P_{ijk}$ for a second spin-$3$ field related to it as $h_{ijk}$ is related to $\Phi_{ijk}$.  This yields a two-spin-$3$-potential (non local) action analogous to the bimetric formulation of \cite{Bunster:2013tc}.

\section*{Acknowledgments} 
 We thank Xavier Bekaert, Nicolas Boulanger and Andrea Campoleoni for very useful discussions. A.L. is Research Fellow at the Belgian F.R.S.-FNRS. This work was partially supported by the ERC through the ``SyDuGraM" Advanced Grant, by IISN - Belgium (conventions 4.4511.06 and 4.4514.08) and by the ``Communaut\'e Fran\c{c}aise de Belgique" through the ARC program.

 \break

\noindent
{\bf \Large{Appendices}}
\appendix

\section{Conventions}
\label{App:Conventions}

We denote symmetrizations and antisymmetrizations respectively with parentheses and brackets.  These operations are of weight one (projectors), e.g., $A_{((ij))} = A_{(ij)}$.

When dealing with tensor fields of mixed Young symmetries, we follow the convention that antisymmetries are manifest.  So, to project on a given Young symmetry type, one first symmetrizes within the rows and then antisymmetrizes within the columns.

The Levi-Civita $\varepsilon_{ab} $ tensor in the internal plane of the prepotentials is such that $\varepsilon_{12} = 1 = - \varepsilon_{21}$.

\section{Complete set of gauge invariant functions}
\label{App:Complete}
\setcounter{equation}{0}

\subsection{Generalities}

Let $\varphi^A$ be some fields invariant under some gauge symmetries,
\be \delta_\xi \varphi^A = k^A_\alpha \xi^\alpha + k^{A i}_\alpha \partial_i \xi^\alpha + k^{Aij}_\alpha \partial_i \partial_j \xi^\alpha \label{AppGauge}
\ee
where for definiteness, we have assumed that the gauge parameters and their derivatives up to second order appear in the gauge transformations.  The discussion would proceed in the same way if there were higher derivatives present in (\ref{AppGauge}).  We also assume that the coefficients $k^A_\alpha$, $k^{A i}_\alpha$ and $k^{Aij}_\alpha$ do not involve the fields, so that the gauge transformations are of zeroth order in the fields (and of course linear in the gauge parameters).

We consider local functions, i.e., functions $f(\varphi^A, \partial_i \varphi^A, \cdots , \partial_{i_1} \partial_{i_2} \cdots \partial_{i_k} \varphi^A)$ of the fields and their derivatives up to some finite but unspecified order.  That unspecified order can depend on $f$.  We denote such local funtions as $f([\varphi^A)])$.  Among the local functions, the gauge invariant ones are particularly important. In our linear theories, non trivial (i.e., not identically constant) local functions that are gauge invariant exist.  For instance, the components of the (linearized) Riemann tensor are local gauge invariant functions under the (linearized) diffeomorphisms.    [Gauge symmetries that involve the fields might not allow for non trivial  local gauge invariant functions.  This occurs for example in the case of full diffeomorphism invariance where even the scalar curvature (say) transforms under change of coordinates, $\delta_\xi  R = \xi^i \partial_i R$ (transport term).]

The local functions are functions on the ``jet spaces'' $J^k$, which can be viewed, in the free theories investigated here, as the vector spaces with coordinates given by the field components $\varphi^A$ and their successive derivatives up to order $k$. The gauge orbits obtained by integrating the gauge transformations are $m$-dimensional planes in those vector spaces $J^k$, where $m$ is the number of independent gauge parameter components and their derivatives (effectively) appearing in the gauge transformations of the fields and their derivatives up to order $k$.  

For instance, for a free spin $3$-field in $3$ dimensions, $J^0$ has dimension $10$ because there are $10$ independent undifferentiated field components $h_{ijk}$.  The gauge orbits have also dimension $10$ since there are $18$ independent derivatives $\partial_{k}\xi_{ij}$ of the gauge parameters  but only $10$ of them, the symmetrized ones $\partial_{(k}\xi_{ij)}$ effectively appear in the gauge transformations. Accordingly, $J^0$ is a single gauge orbit.  Similarly, $J^1$ has dimension $10 + 30 = 40$, the new coordinates being the $30$ derivatives $\partial_m h_{ijk}$ of the fields.  There are $36$ independent second derivatives of the gauge parameters but only $30$ of them effectively act in the gauge transformations of the $\partial_m h_{ijk}$. The jet space $J^1$ reduces again to a single gauge orbit. This is also true for $J^2$. It is only in the jet spaces $J^k$ with $k \geq 3$ that the gauge orbits have a dimension strictly smaller than the dimension of the corresponding jet spaces.  For $J^3$, which has dimension $10$ (number of undifferentiated field components $h_{ijk}$)  $+ 30$ (number of $\partial_m h_{ijk}$) $+ 60$ (number of $\partial_{m}\partial_{n} h_{ijk}$) $ + 100$ (number of $\partial_{m}\partial_{n} \partial_q h_{ijk}$) $ = 200$, the gauge orbits have dimension $10$ (number of effective $\partial_{k}\xi_{ij}$)  $+ 30$ (number of effective $\partial_{k} \partial_m \xi_{ij}$)  $ + 60 $ (number of $\partial_{k} \partial_m \partial_q \xi_{ij}$, which are all effective) $+ 90 $ (number of $\partial_{k} \partial_m \partial_q \partial_r \xi_{ij}$, which are all effective) $= 190$.  Accordingly, the quotient space of $J^3$ by the $190$-dimensional planes generated by the gauge transformations has dimension $10$, which is -- as it should -- the number of independent components of the Riemann tensor, which has Young symmetry
$$\yng(3,3).$$

Without loss of generality, we can assume that the gauge invariant functions vanish when the fields $\varphi^A$ and their derivatives vanish (just substract from $f$ the gauge invariant constant $f(0, 0, \cdots, 0)$). 

A set of gauge invariant functions $\{f_\Delta \}$ is said to form a complete set if any gauge invariant function $f$ can be expressed as a function of the $f_\Delta$, $\delta_\xi f = 0 \Rightarrow f= f (f_\Delta)$.  There might be relations among the $f_\Delta$'s (redundancy) but this will not be of concern to us.   In the linear theories considered here, we can construct complete sets of gauge invariant functions that are linear in the fields and their derivatives. 

Consider a definite jet space $J^k$, with $k$ fixed but arbitrary.  Let $\{f_\Delta \}$ be a complete set of gauge invariant functions. The functions $f^{(k)}_\Delta$ in this complete set that involve derivatives of the fields up to order $k$ are defined in $J^k$.  They provide a coordinate system of the linear quotient space of $J^k$ by the gauge orbits $O_k$ (in case of redundancy, one must take a subset of independent $f^{(k)}_\Delta$).   If this were not the case, one could find a gauge invariant function in $J^k$ not expressible in terms of the functions in the complete set.  The trivial orbit of the pure gauge field configurations is the orbit of $0$, on which the gauge invariant functions have been adjusted to vanish.  It follows from these observations that {\em a set  $\{f_\Delta \}$ of gauge invariant functions is a complete set if and only if the condition $f_\Delta = 0$ implies that the fields are pure gauge.}

\subsection{Spin-$s$ Weyl invariance}
We now turn to the proof that a complete set of invariants for higher spin conformal fields in three dimensions is given by the Cotton tensor and its successive derivatives. As we just shown, this is equivalent to the statement that the vanishing of the Cotton tensor implies that the spin-$s$ field is pure gauge.

To determine a complete set of invariants,  we reformulate  the problem as a problem of cohomogy in the successive jet spaces augmented by new fermionic variables, ``the ghosts", and decompose the successive derivatives in irreducible representations of $GL(3)$.  This approach is standard and has been developed successively in the case of spin-$s$ diffeormorphism invariance for  spin $1$ \cite{Dixon:1991wi,Bandelloni:1986wz,Brandt:1989gy,DuboisViolette:1992ye,Henneaux:1993jn}, spin $2$ \cite{Boulanger:2000rq} and spin $s$ \cite{Bekaert:2005ka}.

Weyl invariance for spin-$2$ was treated in \cite{Boulanger:2001he}.  By the same techniques as those developed in that reference, one first takes care of spin-$s$ diffeomorphism invariance and concludes  that diffeomorphism invariance forces the local functions to be functions of the Riemann tensor and its derivatives, or, what is the same in $D=3$, of the Schouten tensor and its derivatives, $f = f([S])$.  Spin-$s$ Weyl invariance becomes then the condition $\delta_\nu f = 0$ for $\delta_\nu S_{i_1 \cdots i_s} = -\partial_{(i_1 i_2} \nu_{i_3 \cdots i_s)}$ (for convenience, we absorb the factor $\frac{s(s-1)}{2}$  in a redefinition of $\nu$).  Furthermore, neither the Schouten tensor nor the gauge parameter $\nu$ are independent since their divergences are constrained by (\ref{BianchiForSs}) and (\ref{BianchiForNus}).

To investigate the problem of Weyl invariance, we shall first consider the problem $\delta_\nu f = 0$ for $\delta_\nu S_{i_1 \cdots i_s} = -\partial_{(i_1 i_2} \nu_{i_3 \cdots i_s)}$ for unconstrained $S$ and $\nu$. We shall then analyse the implications of the constraints (\ref{BianchiForSs}) and (\ref{BianchiForNus}) on the divergences.

We thus consider the problem of computing the cohomology at ``ghost number" zero of the differential $\gamma$ defined by 
\be
\gamma S_{i_1 \cdots i_s} = \partial_{(i_1 i_2} C_{i_3 \cdots i_s)}, \, \; \; \gamma C_{i_1 \cdots i_{s-2}} = 0 \label{AppGammaForS} \ee
We introduce a derivative degree that gives weight zero to the ghosts and weight two to the Schouten tensor.

At derivative degree $0$, we have only the ghosts in the cohomology, but these are at ghost number one, so there is no cohomology at ghost number zero.  At derivative degree $1$, there is again no cohomology at ghost number zero for a similar reason.  The ghost-number-zero variables (Schouten tensor) appear only in derivative degree $2$ and higher.

At derivative degree $2$,  the second derivatives of the ghosts transform in the representation
\begin{eqnarray}
&&  \overbrace{\yng(8)}^{\text{$s-2$ boxes}} \otimes \yng(2) \nonumber \\
&& = \overbrace{\yng(10)}^{\text{$s$ boxes}} \nonumber \\
&& \oplus \overbrace{\yng(9,1)}^{\text{$s-1$ boxes}} \nonumber \\
&& \oplus  \overbrace{\yng(8,2)}^{\text{$s-2$ boxes}} \nonumber
\end{eqnarray}
while the undifferentiated Schouten tensor components transform in the representation
$$ \overbrace{\yng(10)}^{\text{$s$ boxes}}.$$
It follows that the undifferentiated Schouten tensor components form contractible pairs with the derivatives of the ghosts transforming in the same representation and disappear from the cohomology.   There is no cohomology at ghost number zero.  The same story proceeds in the same way, with the derivatives of the Schouten tensor being all ``eaten" through contractible pairs with the corresponding derivatives of the ghosts and no cohomology at ghost number zero, with non trivial generators at ghost number one left over, up to derivative degree $s$.  There one finds for the ghosts:
\begin{eqnarray}
&&  \overbrace{\yng(8)}^{\text{$s-2$ boxes}} \otimes \overbrace{\yng(10)}^{\text{$s$ boxes}} \nonumber \\
&& = \overbrace{\yng(18)}^{\text{$2s -2$ boxes}} \nonumber \\
&& \oplus \overbrace{\yng(17,1)}^{\text{$2s-3$ boxes}} \nonumber \\
&& \oplus \cdots \nonumber \\
&& \oplus  \overbrace{\yng(11,7)}^{\text{$s+1$ boxes}} \nonumber \\
&& \oplus  \overbrace{\yng(10,8)}^{\text{$s$ boxes}} \label{AppWeylGhost}
\end{eqnarray}
and exactly the same decomposition for the representation in which the derivatives of order $s-2$ of the Schouten tensor transform,
$$ \overbrace{\yng(10)}^{\text{$s$ boxes}} \otimes \overbrace{\yng(8)}^{\text{$s-2$ boxes}}.$$
There is exact matching and the generators of derivatives order $s$ form contractible pairs and do not contribute to the cohomology.

At higher derivative order, it is now some of the derivatives of the Schouten tensor that are unmatched, namely those which contain the Cotton tensor
$$\underbrace{ \overbrace{\yng(10,9)}}^{\text{$s$ boxes}}_{\text{$s-1$ boxes}}$$
since these representations (and only those) cannot arise in the decomposition of the derivatives of the ghosts of order $t>s$ 
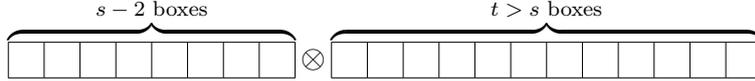
$$\overbrace{\yng(8)}^{\text{$s-2$ boxes}} \otimes \overbrace{\yng(12)}^{\text{$t>s$ boxes}}$$
(the lower line can have at most length $s-2$ as shown by (\ref{AppWeylGhost})).

Accordingly, we can conclude that the $\gamma$-cohomology of the differential defined by (\ref{AppGammaForS}), with unconstrained variables, is given at ghost number zero by the functions $f([C])$ of the Cotton tensor and its derivatives.

We did not take into account so far the constraints (\ref{BianchiForSs}) and (\ref{BianchiForNus}) that the Schouten tensor should obey the Bianchi identity and that the divergence of  the ghost is also determined by its trace.   One must verify that the derivatives of the Schouten tensor that were trivial in the $\gamma$-cohomology because they had an independent ghost partner equal to their $\gamma$-variation, either vanish on account of the constraints or, if they do not vanish, that their ghost partner in the trivial pair also remains different from zero so that both elements in the trivial pair continue being trivial.   

It is easy to convince oneself that this is the case.  The derivatives of the Schouten tensor that remain non-zero after the Bianchi identity has been taken into account may be assumed not to involve a contraction of one derivative index $\partial_i$ with an index of the Schouten tensor, since such terms can be eliminated using the Bianchi identity.  In fact, once we have eliminated such contractions, the remaining derivatives are unconstrained.  A similar situation holds on the ghost side.  If the $\gamma$-variation of a derivative of the Schouten tensor without such contractions involves the ghosts and so is not $\gamma$-closed before the constraints are taken into account, it will remain so after the constraints are taken into account because its $\gamma$-variation necessarily produce independent derivatives of the ghosts without such contractions (in addition to possible terms with such contractions coming from possible traces).

\break

\end{document}